\documentclass[11pt,twocolumn,twoside,a4paper,amsmath,amssymb,aps,showkeys,showpacs]{revtex4}

\usepackage{amsfonts}
\usepackage{graphics} 
\usepackage{epsfig}
\usepackage{fancyheadings}

\begin{document}
\thispagestyle{myheadings}
\rhead[]{}
\lhead[]{}
\chead[Yu.M. Sinyukov, Iu.A. Karpenko]{Non-equilibrium approaches}

\title{Non-equilibrium approaches to the pre-thermal and post-hadronisation stages of A+A collisions}

\author{Yu.M. Sinyukov}
\email{sinyukov@bitp.kiev.ua}

\author{Iu.A. Karpenko}
\email{karpenko@bitp.kiev.ua}

\affiliation{%
Bogolyubov Institute for Theoretical Physics, Metrologichna
14b Kiev 03680 UKRAINE }%

\begin{abstract}
The results related to non-equilibrium phenomena at the very early
and late stages of the processes of A+A collisions are presented. A
good description of the hadron momentum spectra as well as pion and
kaon interferometry data at RHIC is reached within the realistic
dynamical picture of A+A collisions: HydroKinetic Model (HKM). The
model accumulates the following features: not too early
thermalization time; $\tau\geq 1$ fm/c; a developing of the
pre-thermal transverse flows; the effectively more hard, than in the
case of chemical equilibrium, equation of state of expanding
chemically non-equilibrated multi-hadronic gas; a continuous
non-equilibrated emission of hadrons. All these factors lead to a
good description of the mentioned RHIC data, in particular, the
observed $R_{out}/R_{side}$ ratios, solving, therefore, the HBT
puzzle in detailed realistic model.
\end{abstract}



\maketitle

\renewcommand{\thefootnote}{\fnsymbol{footnote}}
{\footnotetext[1]{based on the talks given at XXXIX International
Symposium on Multiparticle Dynamics, Gomel, Belarus September 4-9
2009 and ECT Workshop on flows and dissipation in ultrarelativistic
A+A collisions September 14-18, 2009}

\renewcommand{\thefootnote}{\roman{footnote}}

\section{Introduction}
The typical spacial and temporal scales of the interaction processes
in proton-proton collisions are less or of order 1 fm and 1 fm/c
correspondingly. This is the result of the femtoscopy analysis and
it is agreed with a simple theoretical estimates. The similar
experimental measurements reveals a typical life-time of the system
created in A+A collisions to be at least one order of the magnitude
larger. The femtoscopy, or intensity interferometry measurements
are associated with a set of the points of particle last scattering.
Therefore the above results mean that, even if one tries to consider
the collision of nuclei as a some kind of superposition of the
individual collisions of nucleons of nuclei, the secondary hadrons,
produced in these "elementary" collisions, continue to interact
during the time interval which is much larger than the time-scale
related to the individual nucleon-nucleon collisions. Then one can
conclude that in ultrarelativistic A+A collisions we have to face a
phenomenon of the space-time {\it evolution} of the strongly
interacting matter produced in such processes, and can rise the
question as for a nature of this matter at the early collision
stage. The other crucial points are: whether this matter becomes
thermal in the processes of A+A collisions, and if yes, how does it
evolves and also how are observed particle momentum spectra formed:
in other words, how to describe a particle liberation process, which
gradually destroys the local chemical and thermal equilibrium in
expanding system.

\section{Thermalization and collective flows}
Whether the interaction in the systems formed in A+A collisions  are
strong enough to result in a thermalization and collective effects
such as hydrodynamic flows. A compatibility of the form of pion
rapidity ($y$) spectra with that is predicted by hypothesis of the
longitudinal hydrodynamic flow was stressed first in the pioneer
Landau paper \cite{Landau}. The interferometry signature of the
longitudinal flows, expressed in the specific  $m_T$ and $y$
behavior of the long-radius: $R_{long} \sim
\sqrt{\frac{T}{m_t}}/\cosh{y}$  at high $p_T$ \cite{sin}, was
conformed by NA35/NA49 Collaborations (CERN) \cite {NA35}. The
radial transverse flow in A+A collisions reveal itself in Nu Xu's
plots showing a dependence of the transverse spectra slopes
$T_{eff}$ on hadron masses $m$ (in the non relativistic
approximation $T_{eff} = T_{f.o.} + m\frac{ \langle v^2 \rangle}{2}$
where   $T_{f.o.}$ is the temperature and $\langle v^2 \rangle $ is
the mean squared transverse collective velocity at freeze-out).

The most direct evidence of thermalization and transverse
anisotropic (elliptic) flows at RHIC \cite{elliptic} is related to
the behavior of the so-called $v_2$ coefficients describing the
anisotropy of transverse spectra in non-central A+A collisions. It
is very naturally explained and basically described within
hydrodynamic model for perfect fluid \cite{elliptic1} wherein the
initial geometrical anisotropy transforms into the collective
velocity anisotropy of thermal hadronic gas due to different
gradients of the pressure in out- and in- directions as for the
reaction plane at the initial stage. To reach a quantitative
agreement with the experimental data one needs to use very small
initial time, $\tau \sim 0.5$ fm/c, to start the hydro-evolution. If
it starts at later times, neither the collective velocities, nor
their (and the transverse spectra) anisotropy will be developed to a
sufficient degree.

These results, in fact, brought the two new ideas: first, that the
quark-gluon plasma (QGP), at least at the temperatures not much
higher $T_c$, is the strongly coupled system - sQGP, and so
behaviors as an almost perfect fluid, and, second, that
thermalization happens at very early times of collisions. A
necessity for the thermal pressure to be formed as early as possible
appears also in hydrodynamic description of the central collisions
at RHIC. If one starts the hydrodynamic evolution in "conventional
time" $\tau_i=$1 fm/c {\it without} transverse flow, the latter will
not be developed enough to describe simultaneously the pion, kaon
and proton spectra. The crucial problem is, however, that even the
most optimistic theoretical estimates
give thermalization time 1 - 1.5 fm/c \cite{early}, typically, it is
2 - 3 fm/c. The discrepancy could be even more at LHC energies.

\section{Pre-thermal flows, or Is early thermalization really
needed?}
 The way to solve the problem was proposed in year 2006,
Ref. \cite {sin1} and developed and exploited in Refs. \cite{flow,
JPG}. It was shown that the initial transverse flows in thermal
matter as well as their anisotropy, leading to asymmetry of the
transverse momentum spectra, could be developed at the pre-thermal,
either partonic/string or classical field/Glasma, stage with even
more efficiency than in the case of very early perfect
hydrodynamics. The illustration for the case of partonic free
streaming is presented in Fig. \ref{fig1}.
\begin{figure}
\includegraphics[scale=0.42]{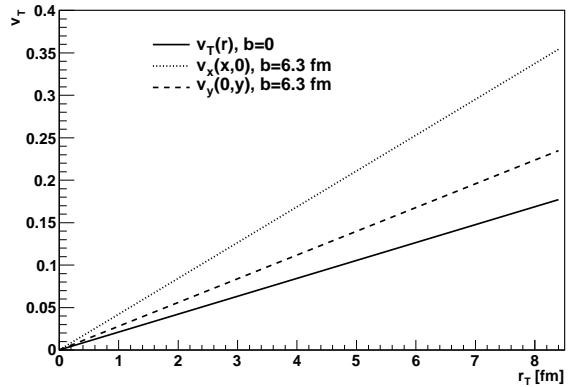}
\caption{\label{fig1}\it The collective velocities developed in
central (b=0) and non-central (b=6.3 fm) Au+Au collisions at the
 pre-thermal stage from proper time $\tau_0=0.3$ fm/c by supposed thermalization time $\tau_i=1$ fm/c
 for scenarios of partonic free streaming in CGC model \cite{JPG}.}
\end{figure}

The results of above mentioned papers show that:

{\it i}) The radial and elliptic flows develop no matter whether a
pressure already established. The general reason for them is an
essential finiteness of the system in transverse direction. Then the
particle number or energy-momentum density flows directed outward
the system cannot be compensated by the inward directed (from
periphery to the center) flows. Just this creates  non-zero net
flows no matter how the collective velocity is defined: according to
Ekkart or to Landau-Lifshitz.

{\it ii}) The specific (linear) dependence of the transverse spectra
slopes on the hadron masses in central A+A collisions and an
anisotropy of the spectra in non-central ones happen  only if, at
least, partial thermalization happens. Then an almost isotropic
particle emission in the local rest frames of the decoupling fluid
elements gains Doppler blue shifts in the Lab system. These shifts
depend on the collective velocities of fluid elements which are
various in different directions in non-central collisions - it leads
to momentum spectra anisotropy.  If no thermalization happen,
similar as at free streaming, the final transverse momentum spectra
will be practically the same as initial transversally isotropic
momentum distribution despite anisotropic collective flows indeed
develop in the non-equilibrated system. In that case also no
specific dependence of the transverse spectra slopes on particle
masses arises.

{\it iii}) So, the results, first published in 2006, show that
whereas the assumption of (partial) thermalization in relativistic A
+ A collisions is really crucial to explain soft physics
observables, the hypotheses of early thermalization  at times less
than 1 fm/c is not necessary.

\section{Phenomenological approach to the pre-thermal evolution}
Of course, the free streaming approximation for the
processes at the very early pre-thermal stage is too rough and leads
to a non local equilibrium structure of the energy-momentum tensor
at the supposed "thermalization" time $\tau_{th}$. In the
forthcoming publication \cite{akksin} we will present in details the
phenomenological approach motivated by Boltzmann equations, which
accounts for the energy and momentum conservation laws and contains
the two parameters: supposed time of thermalization $\tau_{th}$ and
�initial� relaxation time $\tau_{rel}(\tau_0)$. The result is:
if some model or effective QCD theory bring us an energy-momentum
tensor at the very initial time of collision $\tau_0$, then we can
estimate the flows and energy densities at expected time of
thermalization $\tau_{th}$ using equations for the energy-momentum
tensor of ideal fluid with (known) source terms which is associated
with the free evolving initial system when the interaction there is
turned off:
\begin{equation}
\partial_{\mu}\tilde{T}^{\mu \nu}_{hyd}(x)=
- T^{\mu \nu}_{free}(x) \partial_{\mu}{\cal P}_{\tau_0\rightarrow
\tau}(\tau) \label{tensor}
\end{equation} with
\begin{equation}
{\cal P}(\tau)= \left ( \frac{\tau_{f}-\tau}{t_{f}-\tau_{0}}\right
)^{\frac{\tau_{f}-\tau_{0}}
 {\tau_{\tt{rel}}
 (\tau_{0})}}
 \label{P}
\end{equation} and $\tilde{T}^{\mu \nu}_{hyd}$ correspond to $T^{\mu
\nu}_{hyd}$ of ideal fluid with renormalized energy density and
pressure.
$$
\tilde{T}^{\mu \nu}_{hyd}=T^{\mu \nu}_{hyd}(\epsilon \rightarrow
(1-{\cal P}(\tau))\epsilon, p \rightarrow(1-{\cal P}(\tau))p).
$$
Such a method will be included in HydroKinetic Model (HKM)
\cite{PRC, PRL} that accounts for non-equilibrium effects in similar
way; the essence of the model is discussed in what follows.

\section{The evolution in the equilibrated zone and initial
conditions}
At the temperatures higher than the chemical freeze-out
temperature $T_{ch}$ the hydrodynamic evolution is related to
locally equilibrated quark-gluon and hadron phases. The evolution is
described by the conservation law for the energy-momentum tensor of
perfect fluid: $\partial_\nu T^{\mu\nu}(x)=0$ and conservation laws
for baryonic and strange net charge (Q) flows: $\partial_\nu
Q^{\nu}(x)=0$. In Ref. \cite{PRC} the equation of state (EoS) at
$T\geq T_{ch}$ is taken from Ref.\cite{Laine}, that describes well
the QCD lattice data at zero barionic chemical potentials and is
matched with the chemically equilibrated multi-component hadron
resonance gas at $T=175$ MeV. In this work we adjust that EoS to the
parameters at the chemical freeze-out: $\mu_B$ =29 MeV, $\mu_S$ =7
MeV, $\mu_E$ =-1 MeV \cite{Beccatini}.

For very central A+A collisions the initial transverse energy
density profile for such an evolution is supposed to be the Gaussian
as for the variable  $-r^2/R^2$ and with the maximal energy density
$\epsilon(\tau_{th},{\bf r}=0) \equiv \epsilon_0$. We choose the
thermalization time to be $\tau_{th}= 1$ fm/c. The initial
transverse rapidity profile, developed at the pre-thermal stage, is
supposed to be linear in radius $r$: $y_T=\alpha\frac{r}{R}$.  Note
that the parameter $\alpha$ absorbs also a correction for
underestimated resulting transverse flow since in this work we did
not account for the viscosity effects \cite{Teaney} neither at QGP
stage nor at hadronic one. In the longitudinal direction the
boost-invariance with the Bjorken flow is supposed. From the
estimates presented in Ref.\cite{JPG} $R=5.4$ fm for Au+Au
collisions. The only free parameters $\epsilon_0$ and $\alpha$ are
defined from the fitting of the spectra and interferometry radii and
turn out to be $\epsilon_0$= 17 GeV/fm$^3$; $\alpha$=0.35.

\section{The evolution in the non-equilibrated zone and EoS}
At $T<T_{ch}$=165 MeV the system evolves as a chemically and
thermally non-equilibrated hadronic matter. To calculate the EoS we
approximate it below $T_{ch}$ by multi-component hadronic gas where
to guarantee the correct particle number ratios for all the
quasi-stable particles a composition is changed only due to
resonance decays into expanding fluid with possible recombinations.
We include 359 hadron states made of u, d, s quarks with masses up
to 2.6 GeV. Thus, in addition to the equations of energy-momentum
conservation, the equations accounting for the particle number
conservation and resonance decays are added. If one neglects the
thermal motion of heavy resonances, the equations for particle
densities $n_i(x)$ take the form:
\begin{equation}
    \partial_\mu(n_i(x) u^\mu(x))=-\Gamma_i n_i(x) + \sum\limits_j b_{ij}\Gamma_j
    n_j(x)
    \label{paricle_number_conservation}
\end{equation}
where $b_{ij}=B_{ij}N_{ij}$ denote the average number of i-th
particles coming from arbitrary decay of j-th resonance,
$B_{ij}=\Gamma_{ij}/\Gamma_{j,tot}$ is branching ratio, $N_{ij}$ is
a number of i-th particles produced in $j\rightarrow i$ decay
channel. The EoS in this chemically non-equilibrated system depends
now on particle number densities $n_i$ of all the 359 particle
species. So the energy-momentum conservation equations and 359
equations (\ref{paricle_number_conservation}) are solving
simultaneously with calculation of the  EoS,
$p(x)=p(\epsilon(x),\{n_i(x)\})$, at each point $x$. The
thermodynamic region occupied by the EoS is presented in Fig.
\ref{fig2}. Here the different points are related to the various
space-time points of 4-volume swept out by the expanding system
(with initial conditions as is described above).
\begin{figure}
\includegraphics[scale=0.42]{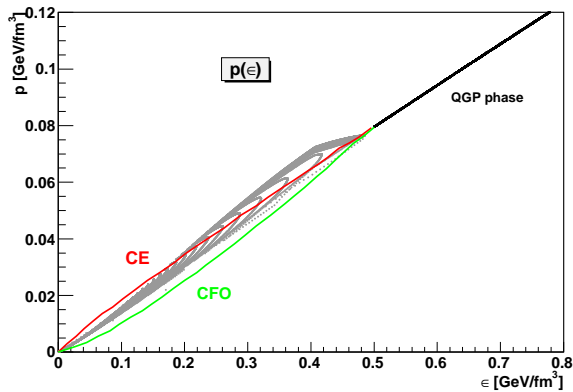}
\caption{\label{fig2}\it The thermodynamic $(\epsilon, p)$-
region occupied by the actual points (grey points) of equation of
state (EoS) $p(x)=p(\epsilon(x),\{n_i(x)\})$ in chemically
non-equilibrated multi-component hadronic gas during its evolution
with IC as described in the body text. The different points are
related to the various space-time points of 4-volume swept out by
the expanding system.  The lines corresponding to chemically
equilibrated (red) and chemically frozen evolution (green) are
marked for a comparison}
\vspace{-0.01\textheight}
\end{figure}

\section{System's decoupling and spectra formation in
HydroKinetic Model - HKM}
During the matter evolution, in fact, at
$T\leq T_{ch}$, hadrons continuously leave the system. Such a
process is described in the HKM by means of the emission function
$S(x,p)$ which is expressed for pions through the term {\it gain},
$G_{\pi}(x,p)$, in Boltzmann equations and
the escape probabilities ${\cal P}_{\pi}(x,p)=\exp(-\int\limits_{t}^{\infty}%
dsR_{\pi+h}(s,{\bf r}+\frac{{\bf p}}{p^0}(s-t),p))$:
$S_{\pi}(x,p)=G_{\pi}(x,p){\cal P}_{\pi}(x,p)$ \cite{PRL,PRC}. For
pion emission in relaxation time approximation $G_{\pi}\approx
f_{\pi}R_{\pi+h}+G_{H\rightarrow\pi}$ where $f_{\pi}(x,p)$ is the
pion phase-space Bose-Einstein distribution, $R_{\pi+h}(x,p)$ is the
total collision rate of the pion, carrying momentum $p$, with all
the hadrons $h$ in the system in a vicinity of point $x$,  the term
$G_{H\rightarrow\pi}$ describes an income of the pions into
phase-space point $(x,p)$ due to the resonance decays. It is
calculated according to the kinematics of decays with simplification
that the spectral function of the resonance $H$ is
$\delta(p^2-\langle m_H\rangle^2)$. The cross-sections in the
hadronic gas, that determine via the collision rate $R_{\pi+h}$ the
escape probabilities ${\cal P}(x,p)$ and emission function $S(x,p)$,
are calculated in accordance with UrQMD method \cite{UrQMD}. The
spectra and correlation function are found from the emission
function $S$ in the standard way (see, e.g., \cite{PRL}).

\section{The results and conclusion}
The systems formed in A+A collisions go through different stages of
evolution: from the initial one, which is far from equilibrium, to
the thermal and chemically equilibrated phases of sQGP and hadronic
matter, then system again becomes non-equilibrated, neither
chemically non thermally. At the late non-equilibrium stage the
medium approximation for  multi-component hadron gas is destroyed
and hadrons are gradually liberated. The correct description of both
non-equilibrium stages: very initial one and the latest stage are
very important since at the first the IC for the fireball evolution
are generated and at the latest one the hadronic momentum spectra
are formed.

It is worthily note that at the first, pre-thermal stage the
collective flows and its anisotropy are developed, as it is
illustrated in Fig. \ref{fig1} for the free streaming approximation. To
go ahead one needs in the correct description of thermalization of
the initially non-equilibrated system of either partons/strings or
Glasma fields. The phenomenological approach (\ref{tensor}) -
(\ref{P}) \cite{akksin} allows one to describe transformation from
an arbitrary initial energy-momentum tensor to the one corresponding
to the perfect fluid in agreement with the conservation laws in the
generalized relaxation time approximation.

The HydroKinetic Model allows one to describe all the stages of the
system evolution as well as a formation of the particle momentum at
the decoupling stage an agreement with the underlying transport
equations. The basic hydrokinetic code, proposed in \cite{PRC}, is
modified now to include decays of resonances into the expanding
hadronic chemically non-equilibrated system. The non-equilibrium
process of the spectra formation is evaluated in the first
approximation when the back reaction of non-equilibrium hadronic
emission on the hydrodynamic evolution is ignored. In formalism of
HKM \cite{PRC} an account of the back reaction of the particle
emission corresponds to an including of the viscosity effects into a
description of hadronic hydro evolution. So here we estimate only
the main characteristics of the dissipative effects in the most
important region corresponding to the maximal pion emission. The
results of our analysis show that for above mentioned initial
conditions for RHIC energies the maximal emission is achieved at the
proper time $\tau \approx 13$ fm/c and transverse radii $r \approx
8.4$ fm. At the corresponding  space-time vicinity, related to the
last stage of the evolution where the spectra are formed, we
estimate $\frac{\eta_{\pi}}{s_{\pi}} = 4.2\frac{1}{4\pi}$,
$\frac{\eta_{p}}{s_{p}} = 0.68 \frac{1}{4\pi}$. The ratio of the
total (over ${\it all}$  hadrons) shear viscosity to total entropy
is $\frac{\eta}{s} = 6.27\frac{1}{4\pi}$. The corresponding mean
free pathes are $\lambda_{\pi} =$ 3 fm, $\lambda_p = $1.2 fm and
Knudsen numbers, defined as the ratio of the m.f.p. to the
hydrodynamic length, are in this region Kn$_{\pi}\approx$ 0.5,
Kn$_{p}\approx$ 0.2. These estimates are important to understand the
evolution of multi-component hadron gas and applicability of
(viscous) hydrodynamic approach to this stage.

\begin{figure*}[!ht]
\vspace{-0.03\textheight}
\begin{tabular*}{1.0\textwidth}{cc}
\begin{tabular}{c}
\includegraphics[scale=0.47]{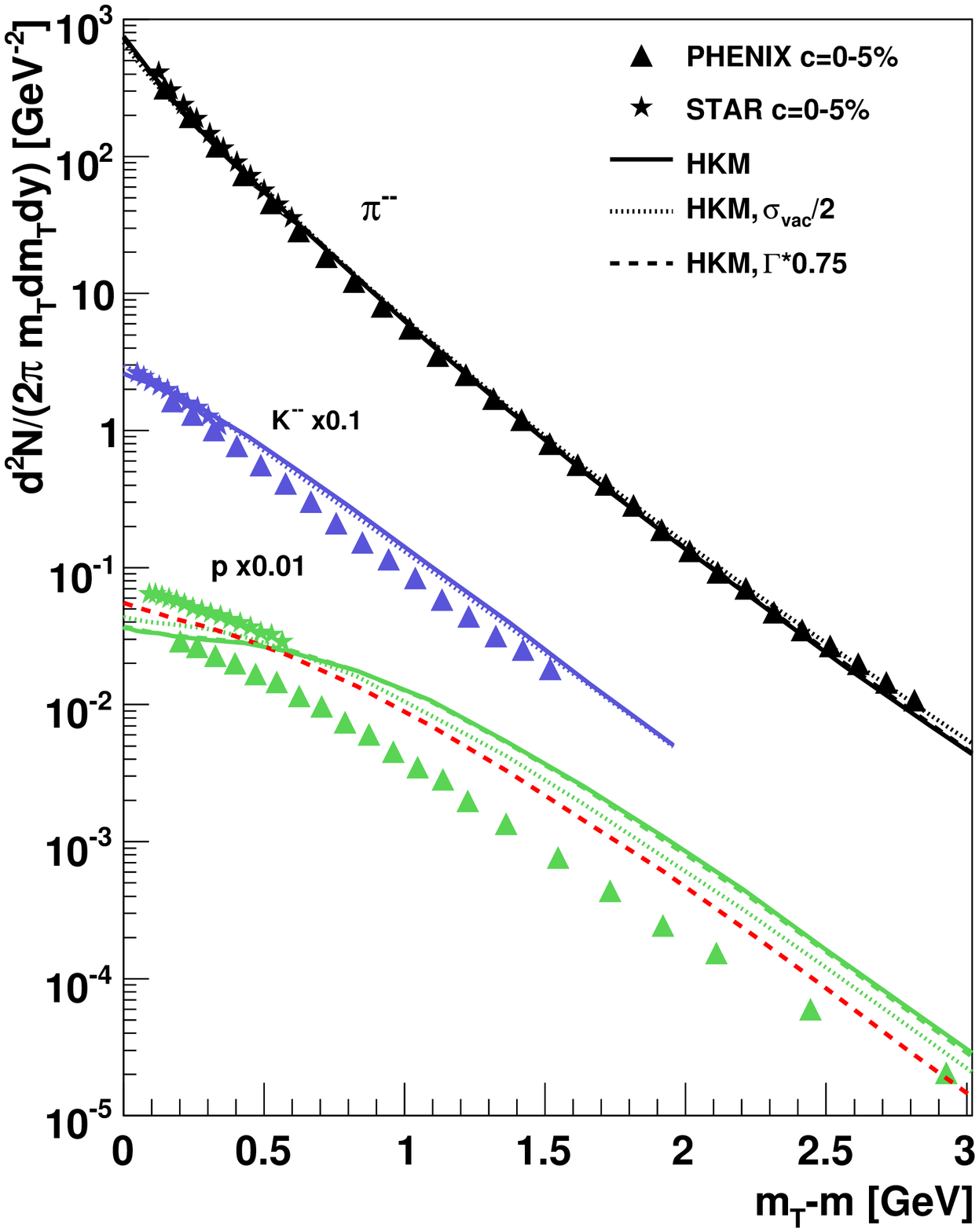} \\
\end{tabular}
&
\begin{tabular}{c}
\includegraphics[scale=0.41]{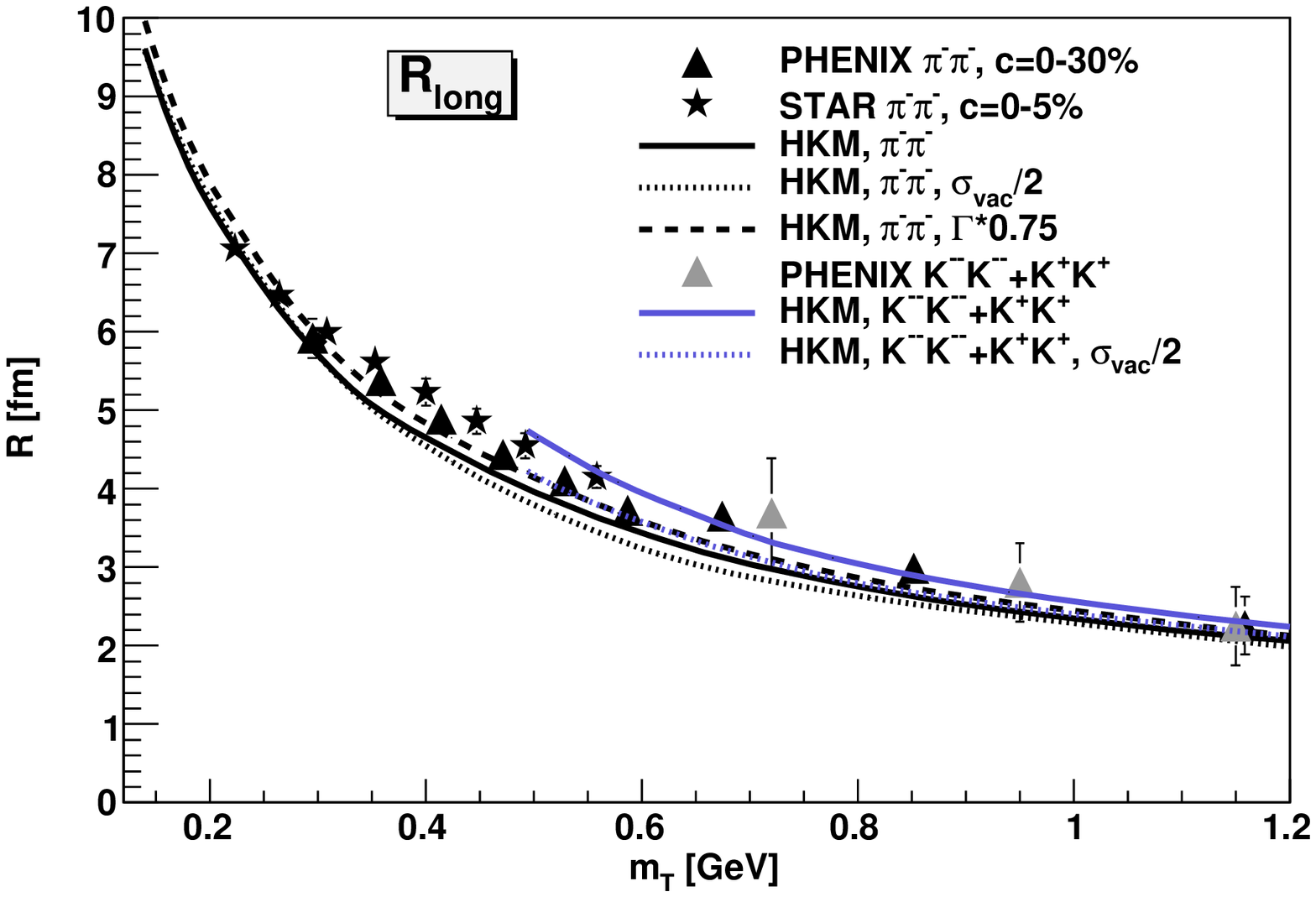} \\
\includegraphics[scale=0.41]{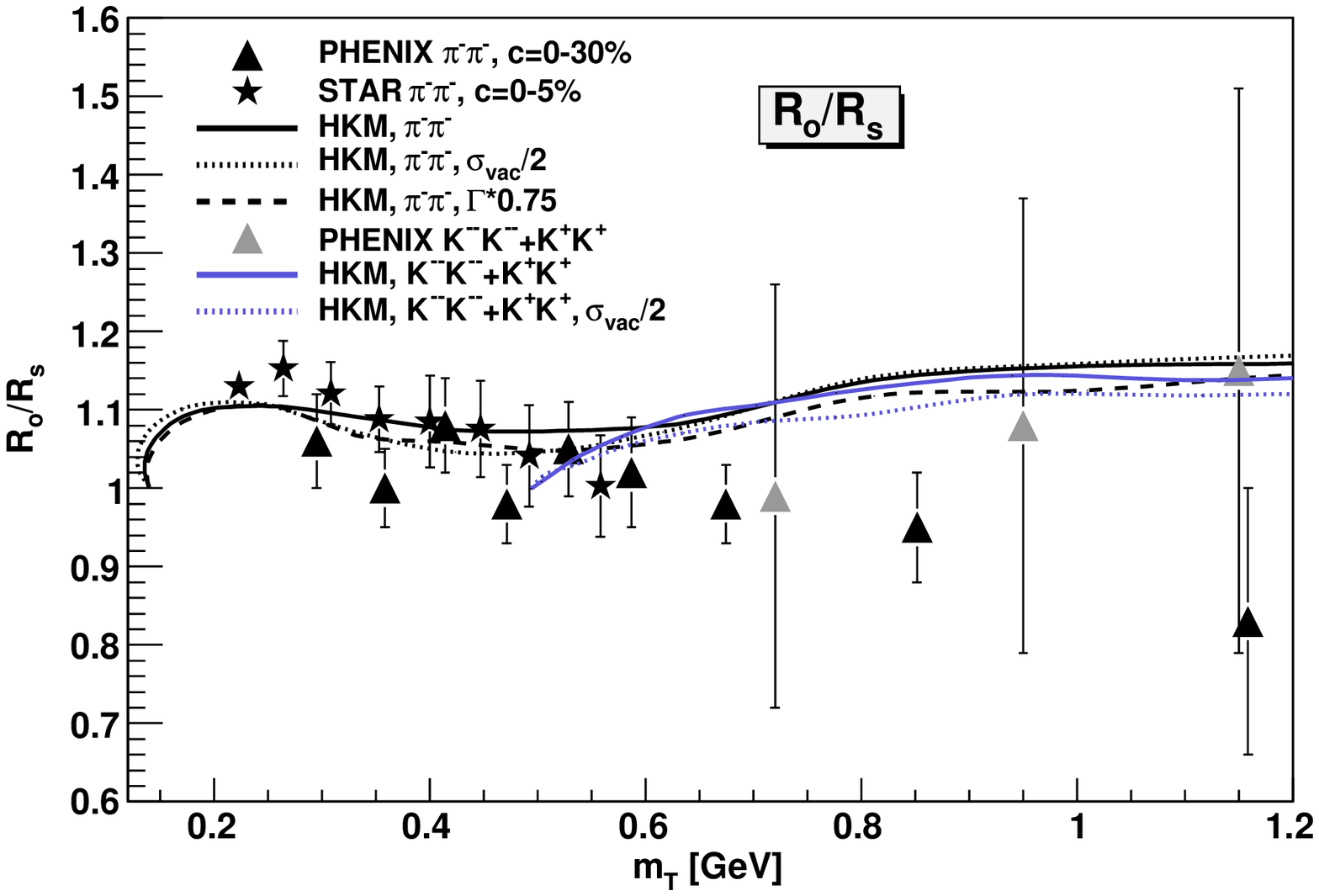}
\end{tabular}
\end{tabular*}
\caption{\label{fig3}{\it Left panel}: The transverse momentum spectra for
negative pions, kaons and protons calculated in HKM for the cases of
the vacuum cross-sections (solid lines), "transport" cross-sections
(dash lines), with account for the resonance recombination processes
(dotted line) and attractive for protons mean field (red line). The
experimental data are taken from STAR \cite{star-spectra} and PHENIX
\cite{phenix-spectra} Collaborations (RHIC BNL).{\it Right panel}:
The HBT radii for the negative pion and negative+positive kaon pairs
calculated in HKM for the same physical conditions as at the left
panel. The experimental data: STAR \cite{star-hbt} and PHENIX
\cite{phenix-hbt, phenix-hbt-kaon}.}
\end{figure*}

The crucial point for any model of space-time evolution is
simultaneous description of hadronic momentum spectra (in absolute
units!), and measured space-time scales: the interferometry, or HBT
radii. The presented approach HKM leads to a good description of the
experimental data as demonstrated in Fig. \ref{fig3}. Because of
lack of the room, we demonstrate only $R_{out}/R_{side}$ ratios, -
our results (not shown)  for the transverse radii $R_{out}$ and
$R_{side}$ are also in good agreement with  the data. As for the
absolute values of the spectra, it is worthy to stress that
parameters of the chemical freeze-out are chosen in accordance with
STAR RHIC data, while for kaons and protons the PHENIX and STAR
results are noticeably different. We analyze an influence of the
main physical factors on the spectra and HBT radii. For this aim we
consider interactions in multi-hadronic gas with different
cross-sections: vacuum ones $\sigma_{ij}$ and with "transport"
cross-sections, $\sigma_{ij}/2$, and also we analyze the possibility
when not only decays of resonances happen but also recombination
processes for non strange resonances takes place, we simulated the
latter effect by utilization of the effective resonance widths  in
Eqs.(\ref{paricle_number_conservation}), $\Gamma_i\rightarrow 0.75
\Gamma_i$, that corresponds to 50 percent increase of effective
life-times of  resonances due to the recombination processes. All
these modifications shift significantly the time distribution of the
hadronic emission but, as it follows from Fig. \ref{fig3}, they do
not influent essentially on pion and kaon spectra and the
interferometry radii. It illustrates and extend the general analytic
results \cite{APSD} that these quantities do not depend
significantly on the freeze-out time at the isentropic evolution. As
for the proton transverse spectrum, its best description requires,
probably, the mean field contribution, which is attractive for
protons \cite{Shuryak} and so leads to reduction of their velocities
in soft momentum region (central part of the fireball). The red dot
line in Fig. \ref{fig3} (left) corresponds to 14 percent reduction
of the proton transverse rapidities in the region $y_T<1$.

Summarizing we would like to emphasis  that correct approaches to
the non-equilibrated stages in A+A collisions, that are utilized
within the HydroKinetic Model lead to a good simultaneous
description of the spectra and space-time scales. The model
accumulates the following features: not too early thermalization
time, $\tau\geq 1$ fm/c; a developing of the pre-thermal transverse
flows; the effectively more hard, than in the case of chemical
equilibrium, equation of state in the chemically non-equilibrated
multi-hadronic gas; a continuous non-equilibrated emission of
hadrons. All these factors improve the description of the
observables at RHIC, in particular, lead to reduction of pion and
kaon $R_{out}/R_{side}$ ratios, solving, therefore, the HBT puzzle
in detailed realistic model.

\section{Acknowledgement}
The research was carried out within the scope of the ERG (GDRE):
Heavy ions at ultrarelativistic energies, Agreement with MESU
F33/461-2009. The work was supported in part by the Program
``Fundamental Properties of Physical Systems under Extreme
Conditions"  of the Bureau of the Section of Physics and Astronomy
of NAS of Ukraine.

\label{last}

\end{document}